\documentclass[osajnl,twocolumn, showpacs, superscriptaddress,10pt]{revtex4-1}   %% LaTeX 2e (preferred)
\usepackage{amsmath,amssymb,graphicx,natbib}
\usepackage{epstopdf}

\begin{document}

\title{A Many-Atom Cavity QED System with Homogeneous Atom-Cavity Coupling}

\author{Jongmin Lee}
\author{Geert Vrijsen}
\author{Igor Teper}
\author{Onur Hosten}
\author{Mark A. Kasevich}\email{Corresponding author: kasevich@stanford.edu}
\affiliation{Physics Department, Stanford University, Stanford, CA 94305, USA}

\date{\today}

\begin{abstract}
We demonstrate a many-atom-cavity system with a high-finesse dual-wavelength standing wave cavity in which all participating rubidium atoms are nearly identically coupled to a 780-nm cavity mode. This homogeneous coupling is enforced by a one-dimensional optical lattice formed by the field of a 1560-nm cavity mode.
\end{abstract}

\ocis{(020.0020) Atomic and molecular physics, (120.3940) Metrology, (140.4780) Optical resonators, (300.6260) Spectroscopy, diode lasers}

\maketitle

There has been growing interest in collective interactions of large ensembles of atoms with cavity fields, in addition to experiments~\cite{MabuchiRev,KimbleRev1} pursuing cavity quantum electrodynamics (cQED) with individual atoms. Topics studied include cavity-aided entanglement generation (spin squeezing) for quantum-enhanced metrology~\cite{Vladan1,Vladan2,Thompson1}; opto-mechanics with atoms, where collective motional degrees of freedom are coupled to cavity fields~\cite{Esslinger1,Purdy10}; cavity-enhanced atomic quantum memories for quantum information processing~\cite{Vladan3}; and ultra-narrow-linewidth lasers using narrow-transition ultra-cold atoms as the gain medium for metrological purposes~\cite{Holland,Thompson2}. Important in many such systems is the inhomogeneity in the coupling strength between the participating atoms and the cavity field, which degrades the coherence of the interactions and complicates the dynamics and the analysis of the basic physics by obscuring the relevant system parameters.~\cite{Teper08}.

Limiting our attention to Fabry-P{\'e}rot type cavities, in which the modes are standing waves, homogeneous coupling of atoms to the relevant cavity field (the probe mode) can be achieved by tightly trapping the atoms with a spatial period that is commensurate with the wavelength of the probe. Thus far, experimentally realized trapping configurations have involved incommensurate trap-probe periods, resulting in inhomogeneous atom-probe couplings that often require the definition of effective, averaged coupling constants ~\cite{Purdy10}. Two recent efforts came to our attention that investigate commensurate dual-wavelength cavity designs; one utilizes a traveling-wave cavity to trap and probe atoms in which there is no particular atom registration~\cite{Simon}, and the other utilizes a standing wave cavity, for the different purpose of investigating atomic self organization~\cite{Kyle}.

\begin{figure}
\centering
\includegraphics[width=0.95\columnwidth]{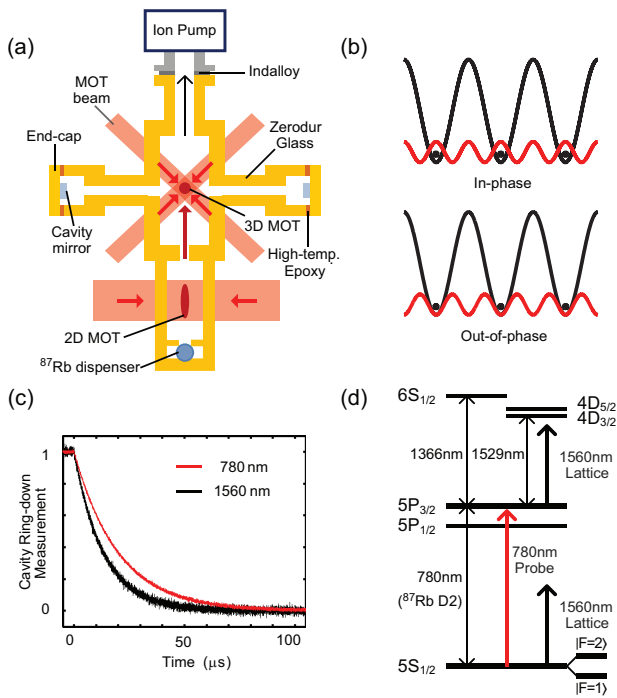}
\caption{(a)~Schematic of the cavity assembly. (b)~In-phase and out-of-phase atom registrations at the cavity center. Black curves: lattice potential due to the $1560\,\mathrm{nm}$ light; red curves: 780\;nm probe light intensity. (c)~Ring-down measurements of the $780\,\mathrm{nm}$ and $1560\,\mathrm{nm}$ cavity modes after an abrupt turn-off of the incident light using an acousto-optic modulator. The measured ring-down times are $\tau_{780} = 19.9$\;$\mu$s and $\tau_{1560} = 13.3$\;$\mu$s. (d)~Energy level diagram of $^{87}$Rb atoms.}
\label{fig1}
\end{figure}

In this Letter, we present the realization and some characteristics of a many-atom cQED system employing probe and trapping modes that are commensurate. We use a dual-wavelength cavity with high finesse at both 780\;nm, used to probe the D$_2$ transition in  $^{87}$Rb, and 1560\;nm, used to trap the atoms in a far-detuned one-dimensional lattice. At the central region of the cavity, depending on the exact wavelength relationship, these commensurate probe and trapping wavelengths allow an in-phase as well as an out-of-phase registration of the atoms (Fig.~\ref{fig1}(b)). In the former case the atoms are localized at the maxima of the probe mode profile, attaining a maximal coupling strength, and in the latter they are localized at the minima, showing that properly positioned atoms can be nearly invisible to light circulating inside the cavity. The work presented here builds on some of our previous results ~\cite{Tuchman06,Long07,Teper08}, but the description is self-contained.

At the the heart of our experimental apparatus is a compact cavity assembly schematically shown in Fig.~\ref{fig1}(a), constructed from Zerodur glass (low helium permeability and small thermal expansion coefficient ($\simeq 10^{-8}$/K)). This assembly includes a current-controlled rubidium dispenser (Alvatec) and a two-dimensional magneto-optical trap (MOT), which prepares a collimated beam of atoms that passes through a 1.5\;mm aperture into the main chamber to load a three-dimensional (3D) MOT.  Extension tubes attached to the main chamber hold the low-loss cavity mirrors. The Zerodur glass contacts, except the mirror end-caps, are held together with a sodium silicate molecular bonding agent, and the glass-metal contacts for the ion pump and the dispenser are sealed with Indalloy. The mirror end-caps, to which the mirror substrates are glued, are attached to the extension tubes with a thin layer of high-temperature epoxy (Epotek 353ND). The epoxy contacts of the chamber can endure a baking temperature of $\sim120^\circ$C, permitting us to reach vacuum pressures lower than $\sim 10^{-10}$\;mbar, maintained by a 5\;L/s ion pump.

The 9.9-cm radius-of-curvature mirrors, coated for high reflectivity at both 780\;nm and 1560\;nm (by Research Electro-Optics), form a 10.73\;cm long near-confocal optical resonator whose modes overlap with the 3D MOT at the center of the chamber. The resonance frequency of the cavity is controlled solely by temperature tuning the thickness of the mirror substrates and spacers. The relevant cavity and atomic parameters are given in Table \ref{Table_cavity_constants}. 

\begin{table}
\caption{\label{Table_cavity_constants}Cavity parameters: Finesse $\mathcal{F}$, mode waist $w_0$, free spectral range $\nu_{\mathrm{fsr}}$, maximum atom-cavity coupling strength $g_0$, cavity HWHM linewidth $\kappa$, atomic HWHM linewidth $\gamma$, single atom cooperativity $g_0^2/2 \kappa \gamma$.}
\begin{center}
\begin{tabular}{ccc}
\hline
{ Parameters } &    { 780\;nm cavity } &    { 1560\;nm cavity} \\
\hline
{ $\mathcal{F}$ } & { $175,000$ } & { 117,000 } \\
{ $w_0$ } & { 111\;$\mu$m} & {157\;$\mu$m} \\
{ $\nu_{\mathrm{fsr}}$ } & \multicolumn{2}{c}{{ 1.3964\;GHz}} \\
{ $g_0$ } & {$2 \pi \times 142$\;kHz} & {$-$} \\
{ $\kappa$ } & {$2 \pi \times 3.99$\;kHz} & { $2 \pi \times 5.98$\;kHz } \\
{ $\gamma$ } & \multicolumn{2}{c}{{ $2 \pi \times 3.03$\;MHz}} \\
{ $g_0^2/2 \kappa \gamma$ } & {0.84} & {$-$} \\
\hline
\end{tabular}
\end{center}
\end{table}

\begin{figure}
\centering
\includegraphics[width=0.95\columnwidth]{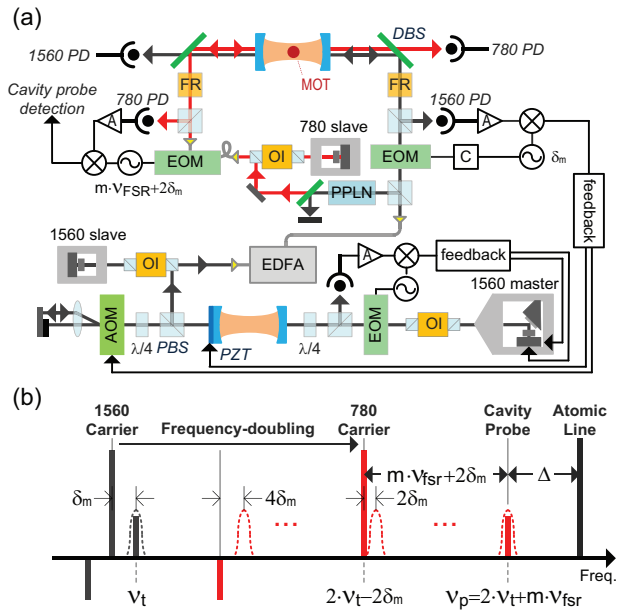}
\caption{Laser-cavity stabilization and generation of spectral lines. (a) The 1560~nm master laser is locked to a reference cavity and the reference cavity is locked to the main cavity. 780~nm light generated by frequency doubling is used to probe the main cavity. EOM, electro-optic modulator; AOM, acousto-optic modulator; PPLN, periodically-poled Lithium Niobate; DBS, dichroic beam splitter; $\lambda/4$, quarter-wave plate; `A', amplifier; `C', controllable attenuator; FR, Faraday rotator; OI, optical isolator; EDFA, erbium doped fiber amplifier. (b) Schematic of the generated optical frequencies with respect to the cavity modes (dotted lines). $\nu_{t}$ is the frequency of the optical lattice; $\nu_{p}$, frequency of the probe; $\delta_{m}$, modulation frequency of the 1560~nm carrier.}
\label{fig2}
\end{figure}

The spectral lines to drive the cavity are generated from an external-cavity diode laser operating at 1560\;nm (New Focus Vortex) stabilized to a TEM00 cavity mode. The schematic for stabilization and generation of spectral lines is shown in Fig.~\ref{fig2}(a). The 1560\;nm master laser is first actively frequency stabilized ($\sim\!5$\;MHz lock bandwidth) to a reference cavity via the Pound-Drever-Hall (PDH) method ~\cite{Drever}. A 1560\;nm slave diode laser is injection-locked to the light transmitted by the reference cavity to remove amplitude fluctuations, and its output is amplified to $\sim\!250$\;mW by an erbium-doped fiber amplifier (EDFA) and split into two paths. The first path gets frequency-doubled to 780\;nm in a periodically-poled lithium niobate (PPLN) waveguide followed by injection into a 780\;nm slave diode laser. The second path is phase-modulated with an electro-optic modulator (EOM) at $\delta_{m}\!\sim\!40$\;MHz, and one of the sidebands is locked to the in-vacuum cavity via the PDH method  ($\sim\!100$\;kHz lock bandwidth), using an acousto-optic modulator (AOM) situated before the 1560\;nm slave diode laser and the piezo-electric transducer of the reference-cavity as the fast and slow feedback elements, respectively. Once locked, the 1560\;nm master laser follows a resonance of the reference cavity, which in turn follows a resonance of the in-vacuum cavity as its length drifts. The frequency-doubled 780\;nm probe light is then also automatically stabilized to the cavity.

The relevant spectral lines are shown in Fig.~\ref{fig2}(b). The 1560\;nm sideband locked to the cavity serves as the trapping lattice light, and a sideband of the 780\;nm slave laser (generated by an EOM at $\sim\!6$\;GHz) serves as the probe light that couples near-resonantly to the trapped atoms, facilitating  the measurement of the cavity resonance frequency in presence of intra-cavity atoms; both of these modes are TEM$_{\rm 00}$. The cavity frequency measurement is accomplished via a heterodyne beatnote measurement between the 780\;nm carrier and sidebands as the probe sideband is frequency swept over the cavity resonance  (similar to PDH method). Not shown in the figures is a microwave horn placed outside of the vacuum chamber that coherently drives the $\sim 6.835$\;GHz $^{87}$Rb hyperfine clock transition of interest ($|F\!=\!1,\, m_\mathrm{f} = 0\rangle\leftrightarrow|F = 2,\, m_\mathrm{f} = 0\rangle$).

In a typical experiment, atoms trapped in the 3D MOT are first further cooled to $\sim 15$\;$\mu$K by switching to a far-detuned MOT ($12 \gamma$ red-detuned) for $\sim 25$\;ms followed by a polarization-gradient cooling stage ($54 \gamma$ red-detuned) for $\sim 7$\;ms. This cooling takes place in the presence of a weak 1560 \;nm optical lattice that is required to keep the lasers locked to the cavity. Near the end of polarization-gradient cooling we adiabatically increase the lattice power by up to a factor of 100 and typically load $10^4-10^5$ atoms distributed over a thousand lattice sites, with a $1/e$ lifetime of 1.1\;s,  mainly limited by background gas collisions. The possibility of applying the regular cooling sequence in the full lattice power condition is hindered by the presence of strong red Stark-shift gradients on the $5P$ excited levels (more than an order of magnitude larger than those experienced by the $5S$ ground states) due to the $5P-4D$ coupling induced by the 1560~nm light (Fig.~\ref{fig1}(d)).

With the onset of the full lattice ($\sim\!40$\;W circulating power), we observe a differential drift between the effective cavity lengths for the 780\;nm and 1560\;nm modes due to heating of the mirror coatings, resulting in a departure of the probe light from resonance by $\sim 100$\;kHz. Nevertheless, equilibrium is attained in a timescale of a hundred milliseconds and can be accounted for in the measurements. We note that even in the absence of light-induced heating, the measured free spectral range (FSR) at 1560\;nm is 7\;kHz smaller than that at 780\;nm. For simplicity, we treat the two FSRs as identical in the rest of the paper; this assumption does not substantively affect the forthcoming discussion.

The $\pi$-polarized probe mode interacts dispersively with the atoms by coupling the $|F = 2\rangle$ ground states to the 5P$_{3/2}$ excited states with an effective frequency detuning $\Delta$, which we typically set around 1\;GHz~$\gg \gamma$. The resulting refractive index of the  $|F = 2\rangle$ atoms shifts the resonance frequency of the probe mode, while the effects of absorption are minor. Note that atoms in $|F = 1\rangle$ states also give rise to probe shifts, but an order of magnitude smaller due to the larger detuning, which we will omit for simplicity. The amount of probe mode shift produced by an atom located at position $\bf{r}$$_{i}$ can be expressed in terms of the position-dependent atom-cavity coupling constant $g_i \equiv g(\bf{r}$$_{i})$, as $\delta \nu_i = g_i^2 / \Delta$. The shift produced by all the atoms is then $\delta \nu = \sum_{i} \delta \nu_i$. We utilize this cavity shift to show that we can trap the atoms entirely either at the anti-nodes or at the nodes of the probe mode, corresponding to in-phase and out-of-phase registrations, respectively.

\begin{figure}
\centering
\includegraphics[width=0.95\columnwidth]{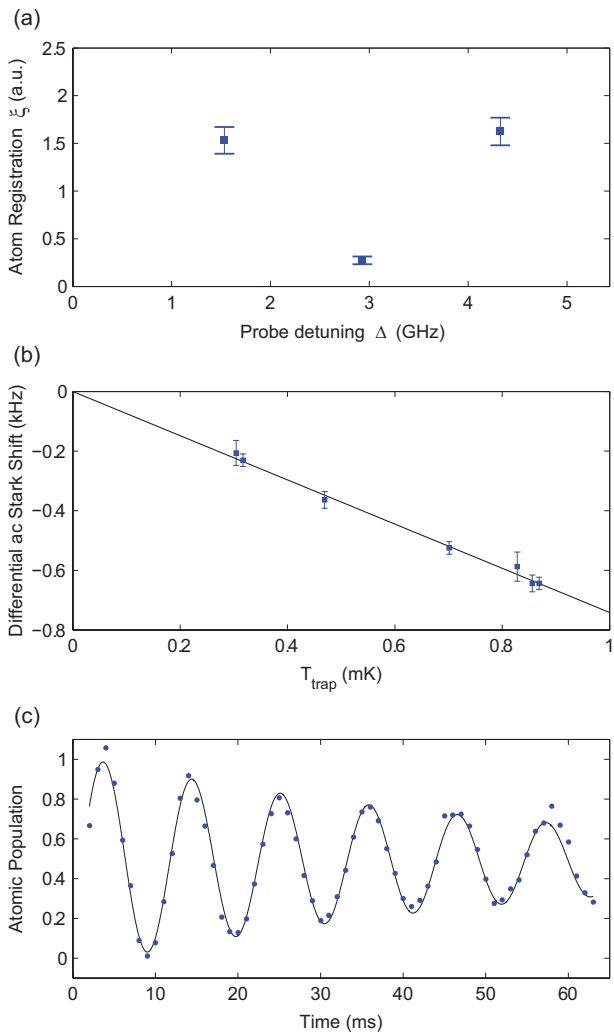}
\caption{(a)~Atom registration parameter as a function of probe detuning from the $F=2$ to $F'=3$ transition. Registration ratio: $6.2\pm0.2$.  (b)~Differential ac Stark shift of the clock states as a function of the estimated trap depth: $-0.74\,\mathrm{Hz/\mu K}$, measured by the microwave transition frequency. (c)~Rabi oscillations with a period of 10.73~ms and 57~ms coherence time measured via the cavity shift.}
\label{fig3}
\end{figure}

Fig.~\ref{fig3}(a) shows the inferred atomic registration parameter $\xi =(\sum_{i}g_i^2)_{\mathrm{trap}}/(\sum_{i}g_i^2)_{\mathrm{mot}}$ for three different probe mode frequencies separated by one FSR. This parameter compares the coupling strength of an unlocalized ensemble to that of a registered ensemble. Here $(\sum_{i}g_i^2)_{\mathrm{mot}}=\delta \nu_\mathrm{mot} \Delta_\mathrm{mot}$, $(\sum_{i}g_i^2)_{\mathrm{trap}}= \delta \nu \Delta$, where $\delta\nu_\mathrm{mot}$ and $\delta\nu$ are the observed cavity shifts before and after the atoms are loaded into the optical lattice, respectively; $\Delta_\mathrm{mot}$ and  $\Delta$  are the effective detunings for the corresponding cases. For every change in the frequency of the probe in steps of a FSR, the wavelength of the probe changes by half a wavelength resulting in the overlap with the lattice near the cavity center to alternate between in-phase and out-of-phase registrations.

The finite temperature of the atomic cloud leads to imperfect registration, as hotter atoms explore larger volumes in the trapping sites, leading especially to a finite coupling for the out-of-phase registration. Thus, the measured ratio of $6.2 \pm 0.2$ of the in-phase to out-of-phase atom registration parameters can provide a direct measure of the temperature of the trapped atoms at a given lattice depth. Assuming a thermal distribution for the atoms in the lattice sites, and utilizing the measured differential ac Stark shift of the hyperfine clock sates (fig.~\ref{fig3}(b)) to infer the lattice depth, we arrive at $\sim\!70$\;$\mu$K for the temperature of the atoms in a lattice depth of 870\;$\mu$K which gives 598\;Hz and  265\;kHz for the transverse and axial trap frequencies. Thus, for this configuration, each atomic ``pancake" in the 1D lattice have rms widths of 22\;$\mu$m in the transverse and 50\;nm in the axial directions, to be compared with the probe waist of 111\;$\mu$m and the probe lattice period of 390 nm respectively. Due to this finite size, the mean coupling strength $\langle g^2 \rangle$ for the in-phase registration is 74\% of the value that would be obtained for perfectly localized atoms, and the coupling strength has a standard deviation of 24\% over the ensemble. However, for atom-probe interaction times much larger than the axial trap period of 3.8\;$\mu$s, the axial inhomogeneity averages out and the relevant standard deviation, e.g. for obtaining probe induced ac Stark shift inhomogeneities, becomes 14\%. For the case of out-of-phase registration, similar calculations show a mean coupling strength of 12\% of the maximal.

The dispersive cavity measurements described here can be used as a sensitive spectroscopic tool. For example, fig.~\ref{fig3}(c) shows Rabi oscillations due to a particular microwave drive resonantly coupling the clock transition. The measured quantity is the population in the upper clock state after a given duration of microwaves, observed by cavity shifts, showing a coherence time of 57\;ms. With similar methods, Ramsey oscillation sequences with echo, implemented inside the lattice ($\pi/2$-pulse, T/2-delay, $\pi$-pulse, T/2-delay, $\pi/2$-pulse), show $1/e$ coherence times up to T=205\;ms.

In summary we demonstrated homogeneous coupling of atoms to a standing wave cavity mode and the potential use of the dispersive measurements as a spectroscopic tool. Although in the described version of the apparatus, the cavity read-out noise is at the atomic shot noise for $10^4$ atoms, with further improvements we expect the possibility of generating spin-squeezed states potentially up to 20\;dB below shot noise variance. Thanks to the homogeneous atom-cavity coupling, the generated squeezed states could be released into free space to be used as input states to atom interferometric sensors for enhanced sensitivity. In addition, the described apparatus might enable atom counting in mesoscopic ensembles with an actual quantized signal, which so far has been challenging~\cite{atomCount}.

\begin{acknowledgements}
This work was funded by DARPA and the MURI on Quantum Metrology sponsored by the Office of Naval Research. We thank the Fejer group for providing the PPLN.
\end{acknowledgements}

%\bibliography{apparatusbib}

\begin{thebibliography}{17}%
\makeatletter
\providecommand \@ifxundefined [1]{%
 \@ifx{#1\undefined}
}%
\providecommand \@ifnum [1]{%
 \ifnum #1\expandafter \@firstoftwo
 \else \expandafter \@secondoftwo
 \fi
}%
\providecommand \@ifx [1]{%
 \ifx #1\expandafter \@firstoftwo
 \else \expandafter \@secondoftwo
 \fi
}%
\providecommand \natexlab [1]{#1}%
\providecommand \enquote  [1]{``#1''}%
\providecommand \bibnamefont  [1]{#1}%
\providecommand \bibfnamefont [1]{#1}%
\providecommand \citenamefont [1]{#1}%
\providecommand \href@noop [0]{\@secondoftwo}%
\providecommand \href [0]{\begingroup \@sanitize@url \@href}%
\providecommand \@href[1]{\@@startlink{#1}\@@href}%
\providecommand \@@href[1]{\endgroup#1\@@endlink}%
\providecommand \@sanitize@url [0]{\catcode `\\12\catcode `\$12\catcode
  `\&12\catcode `\#12\catcode `\^12\catcode `\_12\catcode `\%12\relax}%
\providecommand \@@startlink[1]{}%
\providecommand \@@endlink[0]{}%
\providecommand \url  [0]{\begingroup\@sanitize@url \@url }%
\providecommand \@url [1]{\endgroup\@href {#1}{\urlprefix }}%
\providecommand \urlprefix  [0]{URL }%
\providecommand \Eprint [0]{\href }%
\providecommand \doibase [0]{http://dx.doi.org/}%
\providecommand \selectlanguage [0]{\@gobble}%
\providecommand \bibinfo  [0]{\@secondoftwo}%
\providecommand \bibfield  [0]{\@secondoftwo}%
\providecommand \translation [1]{[#1]}%
\providecommand \BibitemOpen [0]{}%
\providecommand \bibitemStop [0]{}%
\providecommand \bibitemNoStop [0]{.\EOS\space}%
\providecommand \EOS [0]{\spacefactor3000\relax}%
\providecommand \BibitemShut  [1]{\csname bibitem#1\endcsname}%
\let\auto@bib@innerbib\@empty
%</preamble>
\bibitem [{\citenamefont {Mabuchi}\ and\ \citenamefont
  {Doherty}(2002)}]{MabuchiRev}%
  \BibitemOpen
  \bibfield  {author} {\bibinfo {author} {\bibfnamefont {H.}~\bibnamefont
  {Mabuchi}}\ and\ \bibinfo {author} {\bibfnamefont {A.~C.}\ \bibnamefont
  {Doherty}},\ }\href@noop {} {\bibfield  {journal} {\bibinfo  {journal}
  {Science}\ }\textbf {\bibinfo {volume} {298}},\ \bibinfo {pages} {1372}
  (\bibinfo {year} {2002})}\BibitemShut {NoStop}%
\bibitem [{\citenamefont {Kimble}(1998)}]{KimbleRev1}%
  \BibitemOpen
  \bibfield  {author} {\bibinfo {author} {\bibfnamefont {H.~J.}\ \bibnamefont
  {Kimble}},\ }\href@noop {} {\bibfield  {journal} {\bibinfo  {journal}
  {Physica Scripta}\ }\textbf {\bibinfo {volume} {T76}},\ \bibinfo {pages}
  {127} (\bibinfo {year} {1998})}\BibitemShut {NoStop}%
\bibitem [{\citenamefont {Schleier-Smith}\ \emph {et~al.}(2010)\citenamefont
  {Schleier-Smith}, \citenamefont {Leroux},\ and\ \citenamefont
  {Vuleti{\'c}}}]{Vladan1}%
  \BibitemOpen
  \bibfield  {author} {\bibinfo {author} {\bibfnamefont {M.}~\bibnamefont
  {Schleier-Smith}}, \bibinfo {author} {\bibfnamefont {I.~D.}\ \bibnamefont
  {Leroux}}, \ and\ \bibinfo {author} {\bibfnamefont {V.}~\bibnamefont
  {Vuleti{\'c}}},\ }\href@noop {} {\bibfield  {journal} {\bibinfo  {journal}
  {Phys. Rev. Lett.}\ }\textbf {\bibinfo {volume} {104}},\ \bibinfo {pages}
  {073604} (\bibinfo {year} {2010})}\BibitemShut {NoStop}%
\bibitem [{\citenamefont {Leroux}\ \emph {et~al.}(2010)\citenamefont {Leroux},
  \citenamefont {Schleier-Smith},\ and\ \citenamefont {Vuleti{\'c}}}]{Vladan2}%
  \BibitemOpen
  \bibfield  {author} {\bibinfo {author} {\bibfnamefont {I.~D.}\ \bibnamefont
  {Leroux}}, \bibinfo {author} {\bibfnamefont {M.}~\bibnamefont
  {Schleier-Smith}}, \ and\ \bibinfo {author} {\bibfnamefont {V.}~\bibnamefont
  {Vuleti{\'c}}},\ }\href@noop {} {\bibfield  {journal} {\bibinfo  {journal}
  {Phys. Rev. Lett.}\ }\textbf {\bibinfo {volume} {104}},\ \bibinfo {pages}
  {073602} (\bibinfo {year} {2010})}\BibitemShut {NoStop}%
\bibitem [{\citenamefont {Chen}\ \emph {et~al.}(2011)\citenamefont {Chen},
  \citenamefont {Bohnet}, \citenamefont {Sankar}, \citenamefont {Dai},\ and\
  \citenamefont {Thompson}}]{Thompson1}%
  \BibitemOpen
  \bibfield  {author} {\bibinfo {author} {\bibfnamefont {Z.}~\bibnamefont
  {Chen}}, \bibinfo {author} {\bibfnamefont {J.}~\bibnamefont {Bohnet}},
  \bibinfo {author} {\bibfnamefont {S.~R.}\ \bibnamefont {Sankar}}, \bibinfo
  {author} {\bibfnamefont {J.}~\bibnamefont {Dai}}, \ and\ \bibinfo {author}
  {\bibfnamefont {J.~K.}\ \bibnamefont {Thompson}},\ }\href@noop {} {\bibfield
  {journal} {\bibinfo  {journal} {Phys. Rev. Lett.}\ }\textbf {\bibinfo
  {volume} {106}},\ \bibinfo {pages} {133601} (\bibinfo {year}
  {2011})}\BibitemShut {NoStop}%
\bibitem [{\citenamefont {Brennecke}\ \emph {et~al.}(2008)\citenamefont
  {Brennecke}, \citenamefont {Ritter}, \citenamefont {Donner},\ and\
  \citenamefont {Esslinger}}]{Esslinger1}%
  \BibitemOpen
  \bibfield  {author} {\bibinfo {author} {\bibfnamefont {F.}~\bibnamefont
  {Brennecke}}, \bibinfo {author} {\bibfnamefont {S.}~\bibnamefont {Ritter}},
  \bibinfo {author} {\bibfnamefont {T.}~\bibnamefont {Donner}}, \ and\ \bibinfo
  {author} {\bibfnamefont {T.}~\bibnamefont {Esslinger}},\ }\href@noop {}
  {\bibfield  {journal} {\bibinfo  {journal} {Science}\ }\textbf {\bibinfo
  {volume} {322}},\ \bibinfo {pages} {235} (\bibinfo {year}
  {2008})}\BibitemShut {NoStop}%
\bibitem [{\citenamefont {Purdy}\ \emph {et~al.}(2010)\citenamefont {Purdy},
  \citenamefont {Brooks}, \citenamefont {Botter}, \citenamefont {Brahms},
  \citenamefont {Ma},\ and\ \citenamefont {Stamper-Kurn}}]{Purdy10}%
  \BibitemOpen
  \bibfield  {author} {\bibinfo {author} {\bibfnamefont {T.}~\bibnamefont
  {Purdy}}, \bibinfo {author} {\bibfnamefont {D.}~\bibnamefont {Brooks}},
  \bibinfo {author} {\bibfnamefont {T.}~\bibnamefont {Botter}}, \bibinfo
  {author} {\bibfnamefont {N.}~\bibnamefont {Brahms}}, \bibinfo {author}
  {\bibfnamefont {Z.}~\bibnamefont {Ma}}, \ and\ \bibinfo {author}
  {\bibfnamefont {D.}~\bibnamefont {Stamper-Kurn}},\ }\href@noop {} {\bibfield
  {journal} {\bibinfo  {journal} {Phys. Rev. Lett.}\ }\textbf {\bibinfo
  {volume} {105}},\ \bibinfo {pages} {133602} (\bibinfo {year}
  {2010})}\BibitemShut {NoStop}%
\bibitem [{\citenamefont {Tanji}\ \emph {et~al.}(2009)\citenamefont {Tanji},
  \citenamefont {Ghosh}, \citenamefont {Simon}, \citenamefont {Bloom},\ and\
  \citenamefont {Vuleti{\'c}}}]{Vladan3}%
  \BibitemOpen
  \bibfield  {author} {\bibinfo {author} {\bibfnamefont {H.}~\bibnamefont
  {Tanji}}, \bibinfo {author} {\bibfnamefont {S.}~\bibnamefont {Ghosh}},
  \bibinfo {author} {\bibfnamefont {J.}~\bibnamefont {Simon}}, \bibinfo
  {author} {\bibfnamefont {B.}~\bibnamefont {Bloom}}, \ and\ \bibinfo {author}
  {\bibfnamefont {V.}~\bibnamefont {Vuleti{\'c}}},\ }\href@noop {} {\bibfield
  {journal} {\bibinfo  {journal} {Phys. Rev. Lett.}\ }\textbf {\bibinfo
  {volume} {103}},\ \bibinfo {pages} {043601} (\bibinfo {year}
  {2009})}\BibitemShut {NoStop}%
\bibitem [{\citenamefont {Meiser}\ \emph {et~al.}(2009)\citenamefont {Meiser},
  \citenamefont {Ye}, \citenamefont {Carlson},\ and\ \citenamefont
  {Holland}}]{Holland}%
  \BibitemOpen
  \bibfield  {author} {\bibinfo {author} {\bibfnamefont {D.}~\bibnamefont
  {Meiser}}, \bibinfo {author} {\bibfnamefont {J.}~\bibnamefont {Ye}}, \bibinfo
  {author} {\bibfnamefont {D.~R.}\ \bibnamefont {Carlson}}, \ and\ \bibinfo
  {author} {\bibfnamefont {M.}~\bibnamefont {Holland}},\ }\href@noop {}
  {\bibfield  {journal} {\bibinfo  {journal} {Phys. Rev. Lett.}\ }\textbf
  {\bibinfo {volume} {102}},\ \bibinfo {pages} {163601} (\bibinfo {year}
  {2009})}\BibitemShut {NoStop}%
\bibitem [{\citenamefont {Bohnet}\ \emph {et~al.}(2012)\citenamefont {Bohnet},
  \citenamefont {Chen}, \citenamefont {Weiner}, \citenamefont {Meiser},
  \citenamefont {Holland},\ and\ \citenamefont {Thompson}}]{Thompson2}%
  \BibitemOpen
  \bibfield  {author} {\bibinfo {author} {\bibfnamefont {J.~G.}\ \bibnamefont
  {Bohnet}}, \bibinfo {author} {\bibfnamefont {Z.}~\bibnamefont {Chen}},
  \bibinfo {author} {\bibfnamefont {J.~M.}\ \bibnamefont {Weiner}}, \bibinfo
  {author} {\bibfnamefont {D.}~\bibnamefont {Meiser}}, \bibinfo {author}
  {\bibfnamefont {M.~J.}\ \bibnamefont {Holland}}, \ and\ \bibinfo {author}
  {\bibfnamefont {J.~K.}\ \bibnamefont {Thompson}},\ }\href@noop {} {\bibfield
  {journal} {\bibinfo  {journal} {Nature}\ }\textbf {\bibinfo {volume} {484}},\
  \bibinfo {pages} {78} (\bibinfo {year} {2012})}\BibitemShut {NoStop}%
\bibitem [{\citenamefont {Teper}\ \emph {et~al.}(2008)\citenamefont {Teper},
  \citenamefont {Vrijsen}, \citenamefont {Lee},\ and\ \citenamefont
  {Kasevich}}]{Teper08}%
  \BibitemOpen
  \bibfield  {author} {\bibinfo {author} {\bibfnamefont {I.}~\bibnamefont
  {Teper}}, \bibinfo {author} {\bibfnamefont {G.}~\bibnamefont {Vrijsen}},
  \bibinfo {author} {\bibfnamefont {J.}~\bibnamefont {Lee}}, \ and\ \bibinfo
  {author} {\bibfnamefont {M.~A.}\ \bibnamefont {Kasevich}},\ }\href@noop {}
  {\bibfield  {journal} {\bibinfo  {journal} {Phys. Rev. A}\ }\textbf {\bibinfo
  {volume} {78}},\ \bibinfo {pages} {051803} (\bibinfo {year}
  {2008})}\BibitemShut {NoStop}%
\bibitem [{\citenamefont {Bernon}\ \emph {et~al.}(2011)\citenamefont {Bernon},
  \citenamefont {Vanderbruggen}, \citenamefont {Kohlhaas}, \citenamefont
  {Bertoldi}, \citenamefont {Landragin},\ and\ \citenamefont {Bouyer}}]{Simon}%
  \BibitemOpen
  \bibfield  {author} {\bibinfo {author} {\bibfnamefont {S.}~\bibnamefont
  {Bernon}}, \bibinfo {author} {\bibfnamefont {T.}~\bibnamefont
  {Vanderbruggen}}, \bibinfo {author} {\bibfnamefont {R.}~\bibnamefont
  {Kohlhaas}}, \bibinfo {author} {\bibfnamefont {A.}~\bibnamefont {Bertoldi}},
  \bibinfo {author} {\bibfnamefont {A.}~\bibnamefont {Landragin}}, \ and\
  \bibinfo {author} {\bibfnamefont {P.}~\bibnamefont {Bouyer}},\ }\href@noop {}
  {\bibfield  {journal} {\bibinfo  {journal} {New J. Phys.}\ }\textbf {\bibinfo
  {volume} {13}},\ \bibinfo {pages} {065021} (\bibinfo {year}
  {2011})}\BibitemShut {NoStop}%
\bibitem [{\citenamefont {Arnold}\ \emph {et~al.}(2012)\citenamefont {Arnold},
  \citenamefont {Baden},\ and\ \citenamefont {Barrett}}]{Kyle}%
  \BibitemOpen
  \bibfield  {author} {\bibinfo {author} {\bibfnamefont {K.~J.}\ \bibnamefont
  {Arnold}}, \bibinfo {author} {\bibfnamefont {M.~P.}\ \bibnamefont {Baden}}, \
  and\ \bibinfo {author} {\bibfnamefont {M.~D.}\ \bibnamefont {Barrett}},\
  }\href@noop {} {\bibfield  {journal} {\bibinfo  {journal} {Phys. Rev. Lett.}\
  }\textbf {\bibinfo {volume} {109}},\ \bibinfo {pages} {153002} (\bibinfo
  {year} {2012})}\BibitemShut {NoStop}%
\bibitem [{\citenamefont {Tuchman}\ \emph {et~al.}(2006)\citenamefont
  {Tuchman}, \citenamefont {Long}, \citenamefont {Vrijsen}, \citenamefont
  {Boudet}, \citenamefont {Lee},\ and\ \citenamefont {Kasevich}}]{Tuchman06}%
  \BibitemOpen
  \bibfield  {author} {\bibinfo {author} {\bibfnamefont {A.~K.}\ \bibnamefont
  {Tuchman}}, \bibinfo {author} {\bibfnamefont {R.}~\bibnamefont {Long}},
  \bibinfo {author} {\bibfnamefont {G.}~\bibnamefont {Vrijsen}}, \bibinfo
  {author} {\bibfnamefont {J.}~\bibnamefont {Boudet}}, \bibinfo {author}
  {\bibfnamefont {J.}~\bibnamefont {Lee}}, \ and\ \bibinfo {author}
  {\bibfnamefont {M.~A.}\ \bibnamefont {Kasevich}},\ }\href@noop {} {\bibfield
  {journal} {\bibinfo  {journal} {Phys. Rev. A}\ }\textbf {\bibinfo {volume}
  {74}},\ \bibinfo {pages} {053821} (\bibinfo {year} {2006})}\BibitemShut
  {NoStop}%
\bibitem [{\citenamefont {Long}\ \emph {et~al.}(2007)\citenamefont {Long},
  \citenamefont {Tuchman},\ and\ \citenamefont {Kasevich}}]{Long07}%
  \BibitemOpen
  \bibfield  {author} {\bibinfo {author} {\bibfnamefont {R.}~\bibnamefont
  {Long}}, \bibinfo {author} {\bibfnamefont {A.~K.}\ \bibnamefont {Tuchman}}, \
  and\ \bibinfo {author} {\bibfnamefont {M.~A.}\ \bibnamefont {Kasevich}},\
  }\href@noop {} {\bibfield  {journal} {\bibinfo  {journal} {Opt. Lett.}\
  }\textbf {\bibinfo {volume} {32}},\ \bibinfo {pages} {2502} (\bibinfo {year}
  {2007})}\BibitemShut {NoStop}%
\bibitem [{\citenamefont {Drever}\ \emph {et~al.}(1983)\citenamefont {Drever},
  \citenamefont {Hall}, \citenamefont {Kowalski}, \citenamefont {Hough},
  \citenamefont {Ford}, \citenamefont {Munley},\ and\ \citenamefont
  {Ward}}]{Drever}%
  \BibitemOpen
  \bibfield  {author} {\bibinfo {author} {\bibfnamefont {R.~W.~P.}\
  \bibnamefont {Drever}}, \bibinfo {author} {\bibfnamefont {J.~L.}\
  \bibnamefont {Hall}}, \bibinfo {author} {\bibfnamefont {F.~V.}\ \bibnamefont
  {Kowalski}}, \bibinfo {author} {\bibfnamefont {J.}~\bibnamefont {Hough}},
  \bibinfo {author} {\bibfnamefont {G.~M.}\ \bibnamefont {Ford}}, \bibinfo
  {author} {\bibfnamefont {A.~J.}\ \bibnamefont {Munley}}, \ and\ \bibinfo
  {author} {\bibfnamefont {H.}~\bibnamefont {Ward}},\ }\href@noop {} {\bibfield
   {journal} {\bibinfo  {journal} {Appl. Phys. B}\ }\textbf {\bibinfo {volume}
  {31}},\ \bibinfo {pages} {97} (\bibinfo {year} {1983})}\BibitemShut {NoStop}%
\bibitem [{\citenamefont {Zhang}\ \emph {et~al.}(2012)\citenamefont {Zhang},
  \citenamefont {McConnell}, \citenamefont {Cuk}, \citenamefont {Lin},
  \citenamefont {Schleier-Smith}, \citenamefont {Leroux},\ and\ \citenamefont
  {Vuletic}}]{atomCount}%
  \BibitemOpen
  \bibfield  {author} {\bibinfo {author} {\bibfnamefont {H.}~\bibnamefont
  {Zhang}}, \bibinfo {author} {\bibfnamefont {R.}~\bibnamefont {McConnell}},
  \bibinfo {author} {\bibfnamefont {S.}~\bibnamefont {Cuk}}, \bibinfo {author}
  {\bibfnamefont {Q.}~\bibnamefont {Lin}}, \bibinfo {author} {\bibfnamefont
  {M.}~\bibnamefont {Schleier-Smith}}, \bibinfo {author} {\bibfnamefont
  {I.~D.}\ \bibnamefont {Leroux}}, \ and\ \bibinfo {author} {\bibfnamefont
  {V.}~\bibnamefont {Vuletic}},\ }\href@noop {} {\bibfield  {journal} {\bibinfo
   {journal} {Phys. Rev. Lett.}\ }\textbf {\bibinfo {volume} {109}},\ \bibinfo
  {pages} {133603} (\bibinfo {year} {2012})}\BibitemShut {NoStop}%
\end{thebibliography}

%

\end{document}